\documentstyle[aps,preprint, epsfig]{revtex}
\def\beq{\begin{equation}}
\def\eeq{\end{equation}}
\def\be{\begin{eqnarray}}
\def\ee{\end{eqnarray}}
\def\ci{\cite}
\def\bi{\bibitem}

\def\vecq{{\bf q}}

\def\magq{|{\bf q}|}
\def\magp{|{\bf p}|}

\def\yt{{\widetilde y}}
\def\yti{y}
\def\xti{x}

\def\qt{{\widetilde q}}

\begin{document}

\draft

\title{Scaling in many-body systems and proton response}

\author{Omar Benhar}

\address
{
INFN, Sezione Roma 1
\\ Dipartimento di Fisica, Universit\`a ``La Sapienza" \\
Piazzale Aldo Moro, 2. I-00185 Roma, Italy\\
}

\date{\today}
\maketitle

\maketitle


\begin{abstract}

The observation of scaling in processes in which a 
weakly interacting probe delivers large momentum ${\bf q}$ to a 
many-body system reflects the dominance of incoherent scattering 
off target constituents. While a suitably defined scaling function 
can provide rich information on the internal dynamics of the target, 
in general its extraction from the measured cross section requires careful 
consideration of the nature of the interaction driving 
the scattering process. The analysis of deep inelastic electron-proton 
scattering in the target rest frame within standard many-body theory
naturally leads to the emergence of a scaling function that, unlike the 
commonly used structure functions $F_1$ and $F_2$, can be identified 
with the intrinsic proton response. The implications for the theoretical 
analysis of deep inelastic scattering and the interpretation 
of the data are discussed.

\end{abstract}

\pacs{PACS numbers: 13.60.Hb,24.10.Cn,61.12.Bt}


\section{Introduction}
Scaling is observed in a variety of scattering
processes involving many-body systems \ci{west}. For example, at large 
momentum transfer $\magq$ the response of liquid helium measured 
by inclusive scattering of thermal neutrons, which in general depends upon 
{\it both} ${\bf q}$ and the energy transfer $\nu$, exhibits a striking scaling 
behavior, i.e. it becomes a function of the single variable 
$y = (m/\magq)(\nu - \vecq^2/2m)$, $m$ being the mass of the 
helium atom \cite{he}. Scaling in a similar 
variable occurs in inclusive electron-nucleus scattering 
at $\magq >$ 500 MeV and electron energy loss $\nu < Q^2/2M$, 
where $Q^2 = \magq^2  -\nu^2$ and $M$ is the nucleon mass \cite{nuclei}. 
Another most celebrated example is scaling of the deep inelastic proton 
structure functions, measured by lepton scattering at large $Q^2$,
in the Bjorken variable $x=Q^2/2M\nu$ \ci{bj}.

The relation between Bjorken scaling and scaling in the variable $y$, whose
definition and interpretation emerge in a most natural fashion from the
treatment of the scattering process in the target rest frame within 
many-body theory, has been discussed by many authors 
(see, e.g., ref.\ci{west89}).
Recently, deep inelastic data have been also shown to scale
in the variable $\yt = \nu - \magq$ \ci{bps}, related to
{\it both} $y$ {\it and} the Nachtman variable $\xi$ \ci{Nac}, which in 
turn coincides with $x$ in the $Q^2 \rightarrow \infty$ limit.

The observation of scaling in processes driven by different 
interactions clearly indicates that its occurrence reflects 
the dominance of a common reaction mechanism, independent 
of the underlying dynamics. In all instances, scaling is 
indeed a consequence of the 
onset of the kinematical regime in which scattering 
of a weakly interacting probe by a composite target reduces to the 
incoherent sum of elementary scattering processes involving its constituents. 

While the primary goal of scaling analysis is the identification of the 
dominant reaction mechanism, it has to be also emphasized that the scaling 
variable has a straightforward physical interpretation, and a suitably defined 
scaling function, being directly related to the target response, 
contains a great deal of dynamical information. 

In general, extracting the target response from the mesasured 
cross section requires 
careful consideration of the nature of the interaction driving the scattering 
process. In neutron-liquid helium scattering, due to the scalar character
of the probe-constituent coupling, the cross section coincides with
the response up to a kinematical factor \ci{he}. On the other hand, in 
electromagnetic processes the internal dymanics of the target 
and the probe-constituent interaction are nontrivially coupled. 
As a consequence, to obtain the target response in the case of 
electron-nucleus scattering one has to devise a procedure to remove 
from the data the dependence upon the elementary electron-nucleon cross section 
 \ci{nuclei}.
In this paper I will discuss the extension of this method to the 
analysis of deep inelastic scattering (DIS) of electrons by protons.
 
The theoretical treatment of scattering off a many-body system and the 
emergence of scaling as a consequence of the assumptions involved in the 
impulse approximation (IA) are outlined in Section II, where the case of 
neutron-liquid helium scattering is considered as a pedagogical example. 
The procedure leading to identify the target response and 
the scaling variable in the more complex case of electron-nucleus scattering
is traced in Section III, while Section IV describes the application of 
the same procedure to deep inelactic electron-proton scattering,  
and compares the resulting analysis to the standard Bjorken 
scaling analysis. The implications of the proposed approach for 
both the theoretical description of DIS and the interpretation of the data 
are discussed in Section V. Finally, Section VI summarizes the main results 
and states the conclusions.

\section{Scattering off many-body systems in the IA regime and  
$\yti$-scaling}

\subsection{Target response function}

Let us consider scattering off a norelativistic bound system consisting of N 
{\it pointilke} particles of mass $m$, and assume that the 
probe-target interaction be {\it scalar } and {\it weak}, so 
that Born approximation can safely be used. The differential cross section
of the process in which a beam particle, carrying momentum ${\bf k}$ and
energy $E$, is scattered into the solid angle $d\Omega$ with energy 
$E^\prime = E - \nu$ and momentum ${\bf k}^\prime$ can be written 
\beq
\frac{d\sigma}{d\Omega dE^\prime} = \frac{\sigma}{4 \pi}\ 
\frac {|{\bf k}^\prime|}{|{\bf k}|}
S({\bf q},\nu)\ ,
\label{xsec:1}
\eeq
where $\sigma$ is the probe-constituent total cross section. The 
{\it response function} $S({\bf q},\nu)$, containing all the information 
on the structure of the target, is defined as
\be
S({\bf q},\nu) & = & \sum_n | \langle n |\rho_{\bf q}| 0 \rangle |^2
\delta(\nu+E_0-E_n) \label{response:1} \\
&  = & \int \frac{dt}{2\pi}\ {\rm e}^{i\nu t}\
\langle 0 | \rho^\dagger_{\bf q}(t) \rho_{\bf q}(0) | 0 \rangle\ ,
\label{response:2}
\ee
where 
\beq
\rho_{\bf q}(t) = {\rm e}^{iHt}\rho_{\bf q}{\rm e}^{-iHt}\ ,
\label{rho:heis}
\eeq
$H$ is the hamiltonian describing the internal dynamics of the target 
and
\beq
\rho_{\bf q} = \sum_{\bf p} a^\dagger_{{\bf p}+{\bf q}}a_{\bf p}\ ,
\label{rho:0}
\eeq
$a^\dagger_{\bf p}$ and $a_{\bf p}$ being  constituent creation and
annihilation operators, respectively. The target ground and final 
states $| 0 \rangle$ and $| n \rangle$ satisfy the many-body 
Schr\"odinger equations  $H | 0 \rangle=E_0| 0 \rangle$ and 
$H | n \rangle=E_n| n \rangle$. 

Rewriting Eq.(\ref{response:1}) in coordinate space we obtain
\beq
S({\bf q},\nu)= \sum_n \left| \int dR\  \langle n | R \rangle 
\sum_{i=1}^{\rm N} 
{\rm e}^{i {\bf q}\cdot{\bf r}_i}\  \langle R | 0 \rangle \right|^2
\delta(\nu+E_0+E_n)\ ,
\label{resp:coord}
\eeq
where $R\equiv({\bf r}_1,\ldots,{\bf r}_{\rm N})$ specifies the target 
configuration, $\langle R | 0 \rangle$ and
$\langle R | n \rangle$ denote the target initial and final state wave 
functions, respectively,  and the index $i$ labels the struck constituent.
 
The main assumption underlying IA is that, as
the space resolution of a probe delivering momentum $\vecq$ 
is $\sim 1/\magq$, at large enough $\magq$ (typically $\magq \gg 2\pi/d$, 
$d$ being the average separation between target constituents) the 
target is seen 
by the probe as a collection of individual particles.
In addition, final state
interactions (FSI) between the hit constituent, carrying large momentum 
$\sim {\bf q}$, and the residual 
(${\rm N}-1$)-particle system are assumed to be negligibly small.

In the IA regime the scattering process reduces to the 
incoherent sum 
of elementary processes involving only one constituent, the remainig 
(${\rm N}-1$) particles acting as spectators, and Eq.(\ref{resp:coord}) takes 
the simple form
\beq 
S({\bf q},\nu)= \sum_i \sum_n \left| \int dR\ \langle n | R \rangle\  
{\rm e}^{i {\bf q}\cdot{\bf r}_i}\  \langle R | 0 \rangle \right|^2
\delta(\nu+E_0+E_n)\ .
\label{resp:incoh}
\eeq
On account of the absence of FSI, the IA final state $|n\rangle$, carrying total 
momentum ${\bf q}$, exhibits the product structure
\beq
| n \rangle = | {\bf p}^\prime,{\cal R} \rangle = 
|{\bf p}^\prime \rangle \otimes 
| {\cal R}({\bf q} - {\bf p}^\prime) \rangle\ ,
\label{state:IA}
\eeq
its energy being given by
\beq
E_n = E_{{\bf p}^\prime} + E_{{\cal R}}\ .
\label{energy:IA}
\eeq
In the above equations $|{\bf p}^\prime \rangle$ 
denotes the state describing a free particle carrying
momentum ${\bf p}^\prime$ and energy $E_{{\bf p}^\prime}=|{\bf p}^\prime|^2/2m$, 
while $E_{{\cal R}}$ is the energy of the the spectator
(${\rm N}-1$)-particle system, left in the state $| {\cal R} \rangle$ with momentum 
${\bf q} - {\bf p}^\prime$. As a consequence, the sum over final states 
appearing in Eq.(\ref{resp:coord}) can be carried out replacing
\beq
\sum_n | n \rangle \langle n | \rightarrow \int d^3 p^\prime\ 
| {\bf p}^\prime \rangle \langle {\bf p}^\prime |\ \sum_{\cal R}\ 
| {\cal R}({\bf q} - {\bf p}^\prime) \rangle 
\langle {\cal R}({\bf q} - {\bf p}^\prime) |\ . 
\label{sum:IA}
\eeq
Substitution of Eqs.(\ref{state:IA})-(\ref{sum:IA}) 
into Eq.(\ref{resp:incoh}) leads to (see, e.g., ref.\ci{impulse})
\beq
S({\bf q},\nu) = \int d^3 p \int d \epsilon\ 
P({\bf p},\epsilon) \delta(\nu + \epsilon - E_{{\bf p}+{\bf q}})\ ,
\label{resp:IA}
\eeq
where the function
\beq 
P({\bf p},\epsilon) = \sum_{\cal R} \left| 
\langle 0 | {\bf p},{\cal R} \rangle \right|^2 
\delta(\epsilon + E_{\cal R} - E_0)
\label{spec:fcn}
\eeq 
gives the probability of finding a target constituent with momentum 
${\bf p}$ and energy $\epsilon$ in the target ground state.

\subsection{$y$-scaling: neutron scattering off liquid helium 
as an example}

Eqs.(\ref{resp:IA}) and (\ref{spec:fcn}) show that the IA response 
only depends upon ${\bf q}$ and $\nu$ through the energy-conserving 
$\delta$-function, requiring 
\beq
\nu + E_0 - E_{\cal R} - E_{{\bf p}+{\bf q}} = 0\ .
\label{energy:cons}
\eeq
The occurrence of scaling, i.e. the fact that, up to a kinematical factor
$K(\magq,\nu)$, $S({\bf q},\nu)$ becomes a function of a single variable, 
simply reflects the 
fact that in the IA regime, in which energy conservation is expressed by 
Eq.(\ref{energy:cons}), ${\bf q}$ and $\nu$ are no longer 
independent variables. One can then define a new variable
$y=y({\bf q},\nu)$ such that, as $\magq \rightarrow \infty$, 
\beq
K(\magq,\nu) S({\bf q},\nu) \rightarrow F(y)\ .
\label{scal:fcn}
\eeq

All the above results can be immediatly applied to the
case of neutron scattering off liquid helim, in which 
 the energy dependence of $P({\bf p},\epsilon)$ can be safely neglected
 and Eq.(\ref{energy:cons}) takes the form \cite{impulse}
\beq
\nu + \frac{{\bf p}^2}{2 m} - \frac{|{\bf p}+{\bf q}|^2}{2 m} = 0\ ,
\eeq
$m$ being the mass of the helium atom. It follows that, defining 
\beq
y = \frac{m}{\magq} \left( \nu - \frac{\magq^2}{2m} \right)\ ,
\label{he:y}
\eeq
one can rewrite the response in the form 
\beq
S({\bf q},\nu) = \frac{m}{\magq}\ 2 \pi \int_{|y|}^\infty |{\bf p}| d|{\bf p}|\ 
n(|{\bf p}|)\ ,
\label{eq:17}
\eeq
with the constituent momentum distribution defined as
\beq 
n(|{\bf p}|) = \int d\epsilon\ P({\bf p},\epsilon)\ .
\eeq
Eq.(\ref{eq:17}) implies that in the $\magq \rightarrow \infty$ limit  
\beq
\frac{\magq}{m}\  S({\bf q},\nu) \rightarrow F(y)\ ,
\label{he:fy}
\eeq
the kinematical factor being given by $K(\magq,\nu)=
|\partial \nu / \partial p_\parallel|= \magq/m$, with 
$p_\parallel = {\bf p}\cdot{\bf q}/\magq$.

\begin{figure}
\vspace*{1.in}
\centerline
{\epsfig{figure=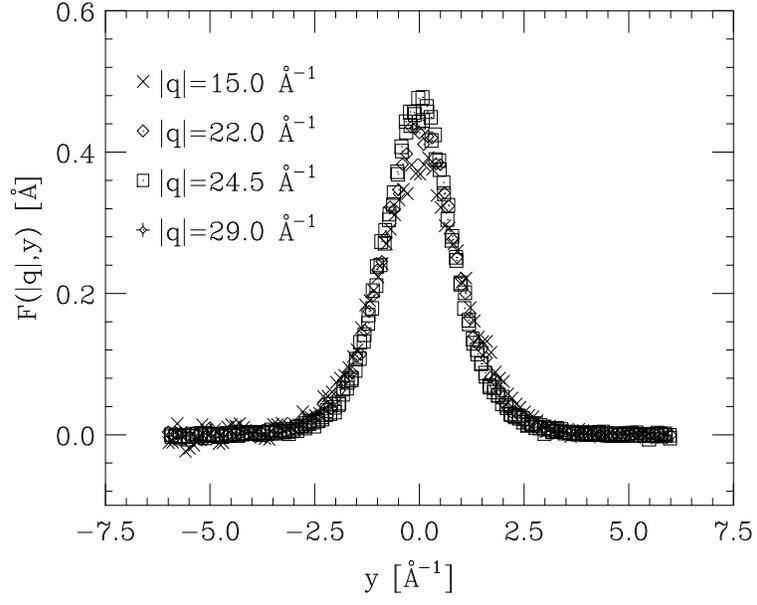
,angle=000,width=10.cm,height=8.0cm}}
\vspace*{.2in}
\caption{
Scaling functions $F(y)$, defined as in Eq.(\protect\ref{he:fy}), measured 
by neutron scattering off superfluid $^4$He at T$=$1.6\ 
$^\circ$K \protect\ci{scal:he}. The data sets are labelled according to the 
magnitude of the momentum transfer $\magq$.
}
\label{helium}
\end{figure}

Fig.\ref{helium} displays the behavior of $F(y)$, defined by the 
above equation, measured in neutron scattering off
superfluid $^4$He at T $=$1.6 $^\circ$K \cite{scal:he}.  
It clearly appears that the curves corresponding to 
$\magq  >$ 15 \AA$^{-1}$ lie on top of one another, 
indicating the onset of the scaling regime.

The variable $y$ defined by Eq.(\ref{he:y}) does have a straightforward 
physical interpretation, as does the scaling function of Eq.(\ref{he:fy}).  
The scaling variable can be identified with the initial longitudinal momentum 
of the struck atom, $p_\parallel$, while $F(y)$ can be related to the momentum 
distribution through
\beq
n(\magp) = -\frac{1}{2\pi}\ \frac{1}{\magp}
\left( \frac{dF}{dy} \right)_{|y|=\magp}\ .  
\label{long:momdis}
\eeq
The relation linking the response in the scaling 
regime to $n(\magp)$ has been been 
extensively exploited to extract momentum distributions of normal and 
superfluid $^4$He from neutron scattering data \ci{he2}. 

\section{Scaling in electron scattering} 

\subsection{Scattering cross section in the IA regime}

The unpolarized electron scattering cross section 
is usually written in the form (see, e.g., ref.\ci{close})
\beq
\frac{d^2 \sigma}{d\Omega dE^\prime} = \frac{\alpha^2}{Q^4}\
\frac{E^\prime}{E}\ L_{\mu \nu} W^{\mu \nu}\ ,
\label{sigmaep:1}
\eeq                            
where $E$ and $E^\prime$ denote the initial and final electron energy, 
respectively, $Q^2=-q^2$, $q = k-k^\prime \equiv (\nu,\vecq)$ being the 
four-momentum transfer,  and $\alpha$ is the fine structure constant. 
The electron tensor $L_{\mu \nu}$ is fully specified by the measured
kinematical variables. In the limit of ultrarelativistic electrons
it reads
\beq
L_{\mu \nu} = 2\left[ k_\mu k_\nu^\prime + k_\nu k_\mu^\prime -
g^{\mu \nu}(k k^\prime) \right]\ . 
\label{Lmunu}
\eeq
The information on the structure of the target is contained 
in the tensor $W_{\mu \nu}$, which can be written in a form reminiscent 
of Eqs.(\ref{response:1}) and (\ref{response:2}) replacing the density 
fluctuation operator $\rho_{\bf q}$ with the electromagnetic 
current $J^\mu$. The resulting expression reads
\be
W^{\mu \nu} & = & \sum_n 
\langle 0 | J^{{\mu^\dagger}} | n \rangle \langle n | J^\nu | 0 \rangle
  \delta^{(4)}(P_0+q-P_n) \label{Wmunu:1} \\
& = & \int \frac{d^4x}{(2\pi)^4}\ {\rm e}^{iqx}\  
\langle 0 | J^{{\mu^\dagger}}(x) J^\nu(0) | 0 \rangle\ ,
\label{Wmunu:2}
\ee
where $P_0 \equiv (M,{\bf 0})$, $M$ being the target mass, and 
$P_n$ denote the four-momenta of the target initial and final states, 
respectively.

In the IA scheme the target tensor of Eqs.(\ref{Wmunu:1}) and 
(\ref{Wmunu:2}) is replaced by a weighted sum of tensors 
describing the electromagnetic structure of target constituents
(compare to Eq.(\ref{resp:IA})):
\beq
W^{\mu\nu} \rightarrow  \sum_i \   
\int d^4 p\ P(p)\ 
w_i^{\mu\nu}({\widetilde p},{\widetilde q})\  
\delta \left( \nu + p_0 - E_{{\bf p}+{\bf q}} \right)\ ,
\label{tensor:IA}
\eeq
where 
$E_{{\bf p}+{\bf q}} = \sqrt{|{\bf p}+{\bf q}|^2 + m^2}$, 
 $m$ being the constituent mass, ${\widetilde p} \equiv (E_{\bf p},{\bf p})$,
 with $E_{\bf p} = \sqrt{|{\bf p}|^2 + m^2 }$, and  
${\widetilde q} \equiv ({\widetilde \nu},{\bf q})$, with 
${\widetilde \nu} = E_{{\bf p}+{\bf q}} - E_{\bf p} = \nu + p_0 
- E_{\bf p}$. The distribution function $P(p)$ 
yields the probability to find a constituent
carrying four-momentum $p\equiv(p_0,{\bf p})$ in the target ground state.  
It has to be emphasized that the tensor $w_i^{\mu\nu}$ describes the 
electromagnetic interaction of the $i$-th target constituent in free space.
Hence, the above equation shows that the IA formalism allows one to 
describe scattering off a {\it bound} constituent in terms of the 
electromagnetic tensor associated with a {\it free} constituent, binding 
effects being taken care of by replacing $p$, $q$ with ${\widetilde p}$,  
 ${\widetilde q}$ (see Appendix A). 

Note that, on account of the replacement 
$q \rightarrow \qt$, $w_i^{\mu \nu}$ is manifestly {\it non} gauge-invariant, 
as $q_\mu w^{\mu \nu} \neq 0$. A somewhat {\it ad hoc} prescription to 
restore gauge invariance, widely used in the theoretical analysis of
electron-nucleus scattering data, has been proposed by De Forest 
back in the 80s \ci{defo} (see Appendix B). More recently, the problem of 
reconciling IA and current conservation in DIS
has been discussed in ref.\ci{franz}. It has 
to be emphasized that, although in general violation of gauge invariance 
is an unpleasant feature inherent in the IA scheme, it does 
not play a role in the $|{\bf q}| \rightarrow \infty$ limit, in which the non
gauge-invariant contributions to the current can be shown to be vanishingly 
small.

Substitution of the IA target tensor (\ref{tensor:IA}) into Eq.(\ref{sigmaep:1}) 
yields
\beq 
\frac{d\sigma}{d\Omega dE^\prime} = \sum_i \  \int d^4 p\ P(p)\ 
\left( \frac{d\sigma_i}{d\Omega dE^\prime} \right)
 \delta \left( \nu + p_0 - E_{{\bf p}+{\bf q}} \right)\ ,
\label{sigmaep:IA}
\eeq
where $(d\sigma_i/d\Omega dE^\prime)$ is the elementary 
electron-constituent cross section, whose structure will be 
discussed in the next Section.

\subsection{Extracting the target response from the cross section}

From Eqs.(\ref{sigmaep:1}), (\ref{tensor:IA}) and (\ref{sigmaep:IA})
it follows that 
\beq
\left( \frac{d\sigma_i}{d\Omega dE^\prime} \right) =
\frac{\alpha^2}{Q^4}\ \frac{E^\prime}{E}\ L_{\mu \nu} 
{\widetilde w}_i^{\mu \nu}\ =
\left( \frac{d\sigma}{d\Omega} \right)_M
\left[ \sigma_{{i_2}} + 2\ \sigma_{{i_1}}\ \tan^2 \frac{\theta}{2} \right]\ ,
\label{sigmac:1}
\eeq 
where $(d\sigma/d\Omega)_M$ denotes the Mott cross section, $\theta$ is
the electron scattering angle and ${\widetilde w}_i^{\mu \nu}$ is the
electromagnetic tensor associated with the $i$-th constituent, corrected 
to restore gauge invariance, whose definition is given in Appendix B. 
 The functions $\sigma_{{i_1}}$ and $\sigma_{{i_2}}$ can be written 
\beq
\sigma_{{i_1}} = \frac{m^2}{E_{{\bf p}}E_{{\bf p}+{\bf q}}}  
\left( w_{i_1} + \frac{1}{2}\ \frac{p_\perp^2}{m^2}\ w_{i_2} \right)
\label{sig:1}
\eeq
and
\beq
\sigma_{i_2} =  \frac{m^2}{E_{{\bf p}}E_{{\bf p}+{\bf q}}} \ 
\frac{q^2}{|{\bf q}|^2}\
 \left\{ w_{i_1} \left( \frac{{\widetilde q}^2}{q^2} - 1 \right) 
+ \frac{w_{i_2}}{m^2} \left[ \frac{q^2}{|{\bf q}|^2}
\left( E_p - {\widetilde \nu}\ \frac{ ({\widetilde q} {\widetilde p}) }
{{\widetilde q}^2} \right)^2  -
\frac{p_\perp^2}{2} \right] \right\}\ ,
\label{sig:2}
\eeq
where $p_\perp$ is the component of the constituent momentum perpendicular 
to the momentum transfer and $w_{i_1}$ and $w_{i_2}$ are the 
{\it constituent structure functions}.

Substitution of Eqs.(\ref{sigmac:1})-(\ref{sig:2}) into Eq.(\ref{sigmaep:IA})
finally leads to the familiar expression of the electron scattering cross section 
in terms of the two {\it target structure functions} $W_1$ and $W_2$: 
\beq
\frac{d^2\sigma}{d\Omega dE^\prime} =  
\left( \frac{d\sigma}{d\Omega} \right)_M \left[ W_2 +
2 W_1 \tan^2 \frac{\theta}{2}  \right]\ ,
\label{sigmaep:2}
\eeq
with
\beq
W_{1,2} = \sum_i \  \int d^4 p\ P(p)\
\sigma_{i_{1,2}}(p,q)\ 
\delta \left( \nu + p_0 - \sqrt{ |{\bf p}+{\bf q}|^2 + m^2} \right)\ .
\label{W12:IA}
\eeq

Eq.(\ref{sigmaep:IA}) shows that in the case of electron 
scattering the target structure function does not appear as a multiplicative 
factor in the cross section. 
In principle, the $p$-dependence of the elementary cross section, coupling 
the internal
 dynamics of the target to the probe-constituent interaction,  
prevents one from extracting the structure function from the data.
In addition, as the integrand in the right hand side of 
Eqs.(\ref{sigmaep:IA}) depends upon $q$ 
through {\it both} $(d\sigma_i/d\Omega dE^\prime)$ {\it and} the argument 
of the energy conserving $\delta$-function, the occurrence of $y$-scaling 
can no longer be established using the simple argument of Section II.

The above problems can be circumvented exploiting the weak 
momentum and energy dependence of the elementary cross section. 
Assuming that the distribution function
$P(p)$ be a rapidly decreasing function, 
the elementary cross section can be evaluated at a constant 
$p={\overline p}$, corresponding to the peak of $P(p)$, and moved out of 
the integral. As a result, one can readily identify the target response 
with the ratio
\beq
S({\bf q},\nu) = \frac{d\sigma}{d\Omega dE^\prime} \left/
 \sum_i 
\left( \frac{d\sigma_i}{d\Omega dE^\prime}  \right)_{p={\overline p}}\ .
\right.
\label{proton:resp}
\eeq
Note that, as $S({\bf q},\nu)$ defined by the above equation 
depends upon $q$ only through the energy conserving delta function, it 
is also expected to exhibit $y$-scaling.

\subsection{$y$-scaling in electron-nucleus scattering}

The procedure described in the previous Section has been extensively employed 
to analyze electron-nucleus ($eA$) scattering data. 
Although in this case the target constituents ($Z$ protons and $N = A - Z$ 
neutrons) are not structureless, $y$-scaling still 
occurs as long as the internal degrees of freedom of the
constituents are not excited by the interaction with the electron probe, 
i.e. as long as the elementary electron-nucleon scattering process is elastic.
In this case, the nucleon structure functions entering Eqs.(\ref{sig:1}) and
 (\ref{sig:2}) can be written ($i = p,n$)
\beq
w_{i_1} = - \frac{ {\widetilde q}^2 }{4m^2}\
\left( F_{{i_1}} + \kappa F_{{i_2}} \right)^2
\label{w1:N}
\eeq
and
\beq
w_{i_2} =
\left[ F^2_{{i_1}} - \frac{ {\widetilde q}^2 }{4m^2} ( \kappa_i F_{{i_2}})^2
 \right]\ ,
\label{w2:N}
\eeq
where
$F_{{i_1}}$ and $F_{{i_2}}$ are the Dirac and Pauli form factors, 
respectively, and $\kappa_i$ denotes the nucleon anomalous magnetic 
moment. 

In analogy with the case of neutron scattering off liquid helium, 
 the requirement of energy conservation provides 
the definition of the scaling variable through
\beq
\nu + M - \sqrt{ (y + |{\bf q}|)^2 + m^2 } - \sqrt{ y^2 + (M-m+B_{0})^2 } = 0 ,
\label{y:N}
\eeq
where $B_0$ is the (positive) minimum nucleon binding energy. The corresponding 
 scaling function is
\beq
F(y) = K({\bf q},y)\ \frac{1}{ (Z {\overline \sigma}_{p} +
N {\overline \sigma}_{n}) } \left( \frac{d\sigma}{d\Omega dE^\prime}
\right)
\label{Fy:N}
\eeq
where $K({\bf q},y) = |\partial \nu / \partial p_\parallel |_{p={\overline p}}$.

\begin{figure}
\vspace*{.5in}
\centerline
{\epsfig{figure=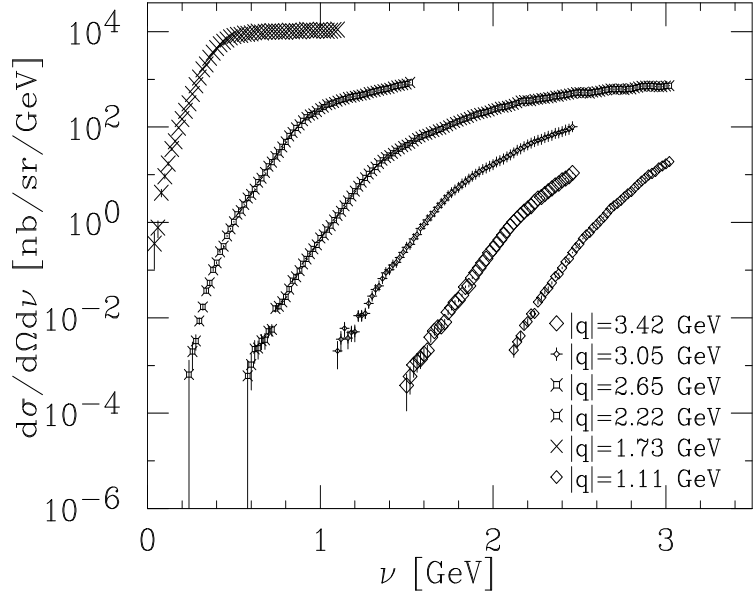
,angle=000,width=10.cm,height=8.0cm}}
\vspace*{.2in}
\caption{
Cross sections for scattering of 4 GeV electrons off Iron, plotted as a 
function of the electron energy loss. The data sets are labelled according 
to the momentum transfer at the quas ielastic peak \protect\ci{E89008}.
}
\label{sigma:eA}
\end{figure}

\begin{figure}
\centerline
{\epsfig{figure=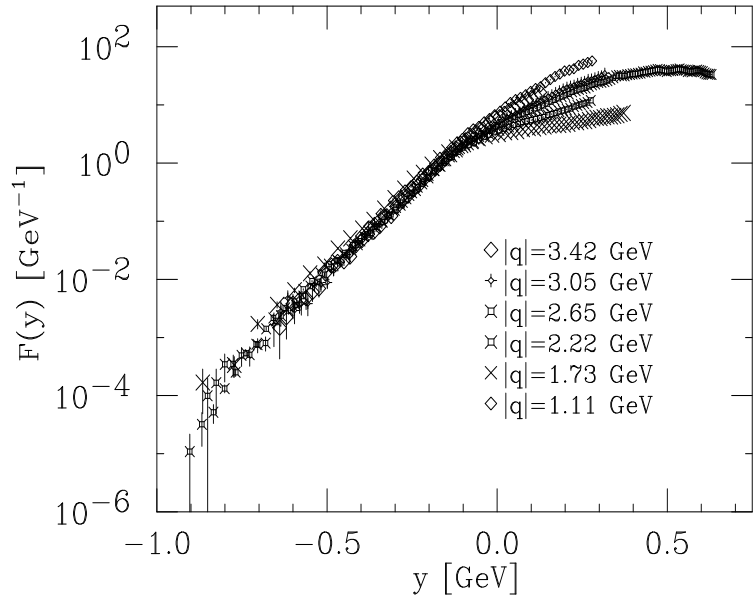
,angle=000,width=10.cm,height=8.0cm}}
\vspace*{.2in}
\caption{
Scaling functions obtained from the definitions of 
Eqs.(\protect\ref{y:N}) and (\protect\ref{Fy:N}) using the 
cross sections shown in fig.\protect\ref{sigma:eA} \protect\ci{E89008}.
}
\label{Fy:eA}
\end{figure}

Fig.\ref{sigma:eA} shows the cross sections recently measured at 
the Thomas Jefferson National Accelerator Facility scattering 4 GeV 
electrons off an Iron target \ci{E89008}. The corresponding scaling 
functions, obtained from the definitions of Eqs.(\ref{y:N}) and (\ref{Fy:N}), 
are shown in fig.\ref{Fy:eA}. It clearly appears that the data, covering 
the range of momentum transfer $ 1 < |{\bf q}| < 4$ GeV, exhibit a striking
scaling behaviour at $y<0$. The scaling violations observed in the 
region of positive $y$, corresponding to large electron energy loss, have to 
be ascribed to the occurrence of inelastic electron-nucleon scattering. 

Although the scaling variable defined by Eq.(\ref{y:N}) can still be 
related to the longitudinal momentum of the struck nucleon, in the case
of $eA$ scattering the constituent binding energy plays a significant 
role, making it more difficult to establish a direct relation between 
$F(y)$ and 
the nucleon momentum distribuution $n(|{\bf p}|)$. A procedure to extract 
$n(|{\bf p}|)$ from $eA$ data in the $y$-scaling 
regime taking into account binding corrections has been developed 
in ref.\ci{cda}.

\section{Scaling in deep inelastic scattering and proton response}
\subsection{The proton as a many-body system}

Let us now make the rather strong assumption that the proton can be viewed 
as a many-body system consisting of {\it bound} pointlike Dirac particles
of mass $m$ and charge $e_i$ (in units of the magnitude of the electron charge), 
and extend the analysis described in the previous Sections 
to deep inelastic electron-proton ($ep$) scattering. In this scenario, as 
in the standard parton model of DIS \ci{close}, one assumes 
that over the short spacetime scale relevant to the scattering process 
confinement does not play a role, so that proton constituents can be described 
in terms of physical states. It has to be emphasized, however, that, unlike 
the parton model, the present approach {\it does not} involve the
additional assumption that proton constituents be on mass shell.

The structure functions appropriate for the case of pointlike
constituents are (compare to Eqs.(\ref{w1:N}) and (\ref{w2:N}))
\beq
w_{i_1} = - e_i^2\ \frac{ {\widetilde q}^2 }{4m^2}\
\label{w1:Q}
\eeq
and
\beq
w_{{i_2}} = e_i^2\ .
\label{w2:Q} 
\eeq
 
The requirement of energy conservation, expressed by the equation
\beq
\nu + p_0 - \sqrt{|{\bf p}+{\bf q}|^2 + m^2} =
\nu + p_0 - p_\parallel - \magq + {\cal O}\left(\frac{1}{\magq}\right) = 0\ ,
\label{en:cons}
\eeq
where $p_0 = M - E_{\cal R}$ and $E_{\cal R} = \sqrt{\magp^2 + M_{\cal R}^2}$, 
$M_{\cal R}$ being the mass of the spectator system, implies that, as 
$\magq \rightarrow \infty$, the quantity
\beq
\yt = \nu - \magq = p_\parallel - p_0
\label{def:yt}
\eeq
becomes independent of ${\bf q}$. Hence, in this limit $S({\bf q},\nu)$, 
defined as in eq.(\ref{proton:resp}), is expected to exhibit scaling in the 
variable $\yt$ \cite{bps}, i.e.
\beq
S({\bf q},\nu) \rightarrow F(\yt). 
\label{F:DIS}
\eeq
Note that in this case the kinematical factor entering the definition of 
the scaling function is $|\partial \nu / \partial p_\parallel | \equiv 1$.

It has to be pointed out that $\yt$ {\it does not} have the same physical 
interpretation as the variable $y$ discussed in the previous sections: 
 it {\it does not} coincide with the constituent 
longitudinal momentum. 
However, as $p_0$ is independent of $q$, scaling in $y = p_\parallel$ 
necessarily implies scaling 
in $\yt$, and {\it viceversa}. The motivation for choosing $\yt$ as scaling 
variable in DIS will be discussed in the next Section.

\subsection{$\yt$-scaling analysys of DIS data}

According to the IA picture, the $\yt$-scaling function can be obtained 
dividing 
either structure function by the appropriate contribution to the elementary
cross section. In fact, from Eqs.(\ref{sigmaep:2})-(\ref{proton:resp}) and 
(\ref{F:DIS}) it follows that, in the $\magq \rightarrow \infty$ limit, 
\beq
S({\bf q},\nu) \ = \ \frac{W_1}{{\overline \sigma}_1} 
\ = \ \frac{W_2}{{\overline \sigma}_2} \rightarrow F(\yt)  \ ,
\eeq
where
\beq
{\overline \sigma}_{1,2} = \sum_i \  
( {\overline \sigma}_{i_{1,2}} )_{p={\overline p}}\ , 
\eeq
with ${\overline p}\equiv({\overline p}_0,{\bf p}_{min})$, 
${\bf p}_{min}$ being the minimum constituent  momentum allowed
in the kinematics specified by ${\bf q}$ and $\nu$. The magnitude 
of ${\bf p}_{min}$ is given by
\beq
|{\bf p}_{min}| = \frac{1}{2} \left|
\frac{M_{\cal R}^2 - (\yt + M)^2}{\yt + M} \right|\ ,
\label{p:min}
\eeq
while ${\overline p}_0$, related to the mass of the spectator system
through
\beq
{\overline p}_0 = M - \sqrt{ |{\bf p}_{min}|^2 + M_{\cal R}^2 }\ ,
\label{p0:bar}
\eeq
can be simply parametrized in terms of the positive quantity $B_0$ 
according to
\beq 
M_{\cal R} = M - m + B_0\ .
\label{recoiling:mass}
\eeq

The above discussion shows that the scaling analysis in terms of 
$\yt$ involves two parameters: the constituent mass, $m$, and $B_0$, 
playing the role of the constituent binding energy of many-body theory.

\begin{figure}
\vspace*{1.in}
\centerline
{\epsfig{figure=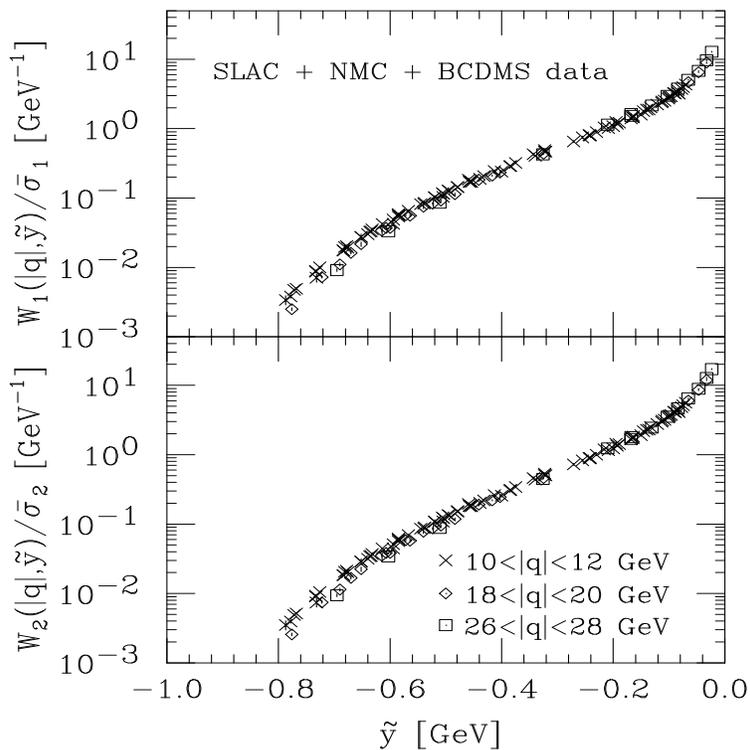,
angle=000,width=10.cm,height=10.0cm}}
\vspace*{.2in}
\caption{
Scaling functions obtained from $W_1/{\overline \sigma}_1$ (upper panel)
and $W_2/{\overline \sigma}_2$ (lower panel). The experimental 
structure functions are taken from refs.\protect\ci{SLAC,NMC,BCDMS} and 
labelled according to the values of $\magq$, whereas
the elementary cross sections have been evaluated 
using Eqs.(\protect\ref{sig:1})-(\protect\ref{sig:2}) and
(\protect\ref{w1:Q})-(\protect\ref{w2:Q}), with $m = 300$ 
MeV and $B_0 = 200$ MeV.
}
\label{DIS:1}
\end{figure}

\begin{figure}
\vspace*{.5in}
\centerline
{\epsfig{figure=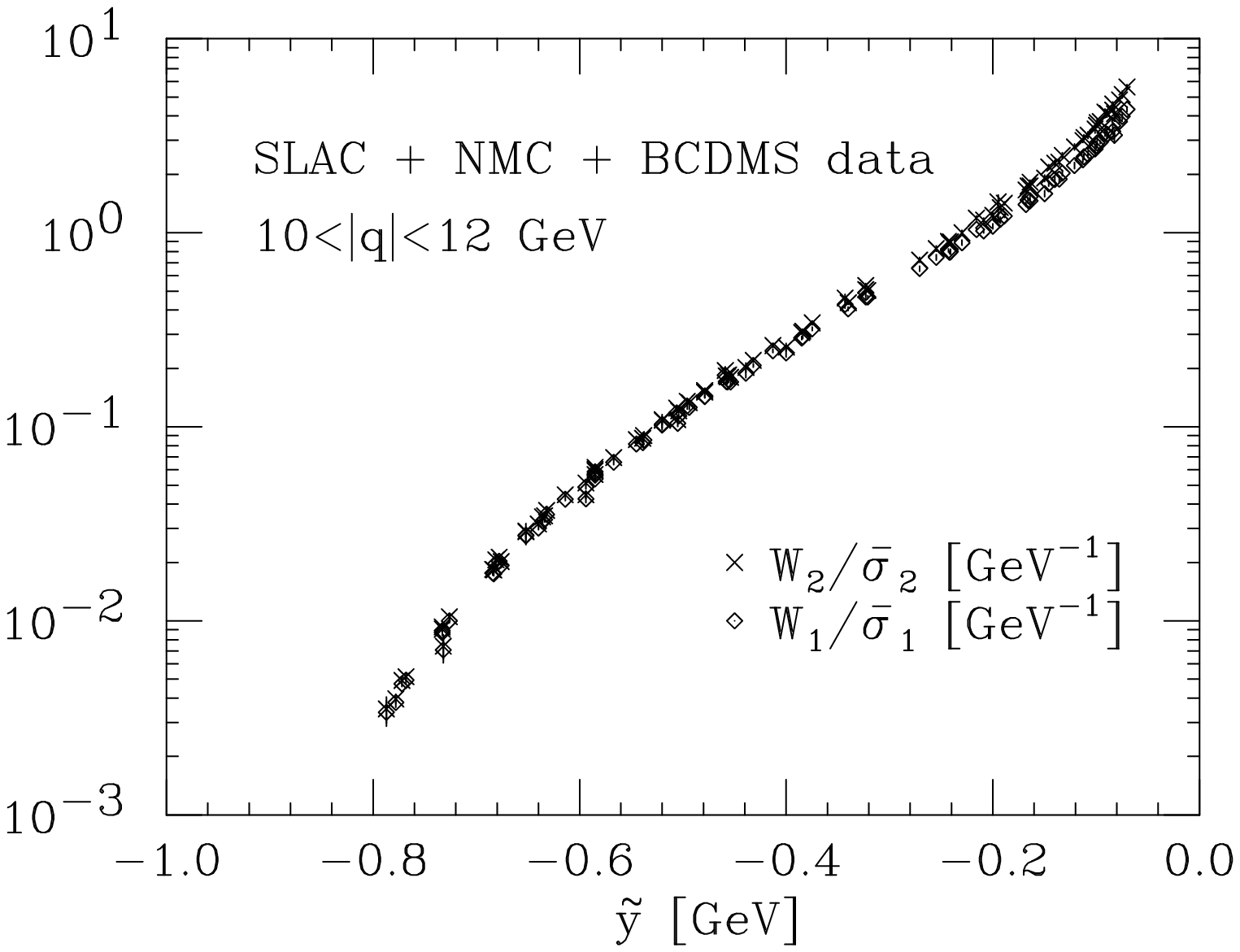,
angle=000,width=10.cm,height=8.0cm}}
\vspace*{.2in}
\caption{
Comparison between $W_1/{\overline \sigma}_1$ (diamonds) and 
$W_2/{\overline \sigma}_2$ (crosses) of fig. \protect\ref{DIS:1} 
at 10 $\leq \magq \leq$ 12 GeV. 
}
\label{DIS:2}
\end{figure}

Fig.\ref{DIS:1} shows the quantities $W_1/{\overline \sigma}_1$ 
(upper panel) and $W_2/{\overline \sigma}_2$ (lower panel),  
obtained from DIS data taken at SLAC \ci{SLAC} and CERN \ci{NMC,BCDMS} and
rearranged in bins of constant $\magq$ centered at 11, 19 and 27 GeV, 
plotted as a function of $\yt$. The structure functions $W_1$ have been
obtained from the tabulated $F_2=\nu W_2$ using the parametrization of 
$R=\sigma_L/\sigma_T$ of 
ref. \ci{RLT}, while the elementary cross sections have been evaluated
from Eqs.(\ref{sig:1})-(\ref{sig:2}) and (\ref{w1:Q})-(\ref{w2:Q}), 
with $m = 300$ MeV and 
$B_0 = 200$ MeV. It clearly appears that in both cases scaling sets in at 
$\magq >$ 10 GeV. The second feature predicted by the IA analysis, i.e. that 
$W_1/{\overline \sigma}_1$ and $W_2/{\overline \sigma}_2$ scale to 
the {\it same} function $F(\yt)$, is illustrated in fig.\ref{DIS:2}.

It has to be emphasized that the occurrence of $\yt$ scaling {\it does not} 
depend upon the values of either $m$ or $B_0$. In particular, it does not 
require that the constituent mass be negligibly small, nor that the 
struck constituent be on mass shell. Varying $m$ and $B_0$ in the range 
$10-300$ MeV does not appreciably affect the scaling behavior displayed in 
figs.\ref{DIS:1} and \ref{DIS:2}. However, it {\it does} affect the scaling 
function extracted from the data. 

\begin{figure}
\centerline
{\epsfig{figure=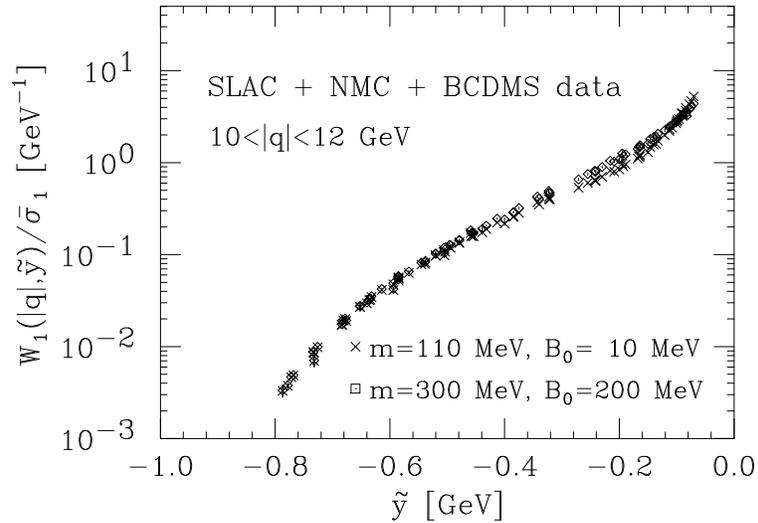,
angle=000,width=10.cm,height=7.0cm}}
\vspace*{.2in}
\caption{
$\yt$-scaling function obtained from the structure functions 
of refs.\protect\ci{SLAC,NMC,BCDMS} corresponding to 
10 $\leq \magq \leq$ 12 GeV. 
Crosses and diamonds represent the results obtained using 
$m = 110$ MeV, $B_0 = 10$ MeV and $m = 300$ MeV, $B_0 = 200$ MeV, respectively.
}
\label{DIS:3}
\end{figure}

While increasing $m$ at fixed $B_0$ leads
to a shift of $F({\widetilde y})$ towards lower ${\widetilde y}$ for
${\widetilde y} > -.5$ GeV,  increasing $B_0$ at fixed constituent mass
produces a shift in the opposite direction.  As a result, the scaling function 
turns out 
to be only sensitive to the difference $m-B_0$, i.e. to the mass of the 
recoiling spectator system $M_{\cal R}$. This feature is illustrated 
in fig.Fig.\ref{DIS:3}, showing that the scaling functions corresponding to 
$m = 110$ MeV, $B_0 = 10$ MeV and $m = 300$ MeV, $B_0 = 200$ MeV, 
yielding the same value of $M_{\cal R}$, lie on top of one another.

\subsection{Comparison between $\yt$- and $\xti$-scaling analysis}

DIS data are usually analyzed in terms of the two dimensionless 
structure functions $F_1 = M W_1$ and $F_2 = \nu W_2$. In the Bjorken 
limit $Q^2, \nu \rightarrow \infty$, with $Q^2/\nu$ finite 
and $\nu/\magq \rightarrow 1$, both $F_1$ 
and $F_2$ exhibit scaling, i.e. they become functions of 
a single variable $x=Q^2/2M\nu$. 

The first issue to be addressed to establish a connection 
between $x$- and $\yt$-scaling is the relation between the 
scaling variables. As pointed out in 
ref.\ci{bps}, $\yt$ is trivially related to another variable commonly used
in the context of DIS, the Nachtman variable $\xi$ \ci{Nac},
through
\beq
- \frac{\yt}{M} = \xi = \frac{2x}{1+\sqrt{1+4M^2x^2/Q^2}}\ ,
\label{yt:csi}
\eeq
where $x$ is the Bjorken variable. From the above equation it 
follows that in the $Q^2 \rightarrow \infty$ limit $\yt$ coincides with $x$ 
up to a constant factor. Comparison between $x$ and $-\yt/M$ at constant 
$\magq = 20\ {\rm MeV}$ shows that the difference is in fact less
then 2 \% over the whole $0 \leq x \leq 1$ range.

The correspondence between $\yt$ and $\xi$ makes it 
clear why $\yt$ is better suited than $y$ for the analysis of DIS.
In addition, it has to be pointed out that the IA scheme provides a simple
physical intepretation of Nachtman's variable, whose definition was
originally obtained in a totally different fashion \ci{Nac}.

\begin{figure}
\vspace*{1.in}
\centerline
{\epsfig{figure=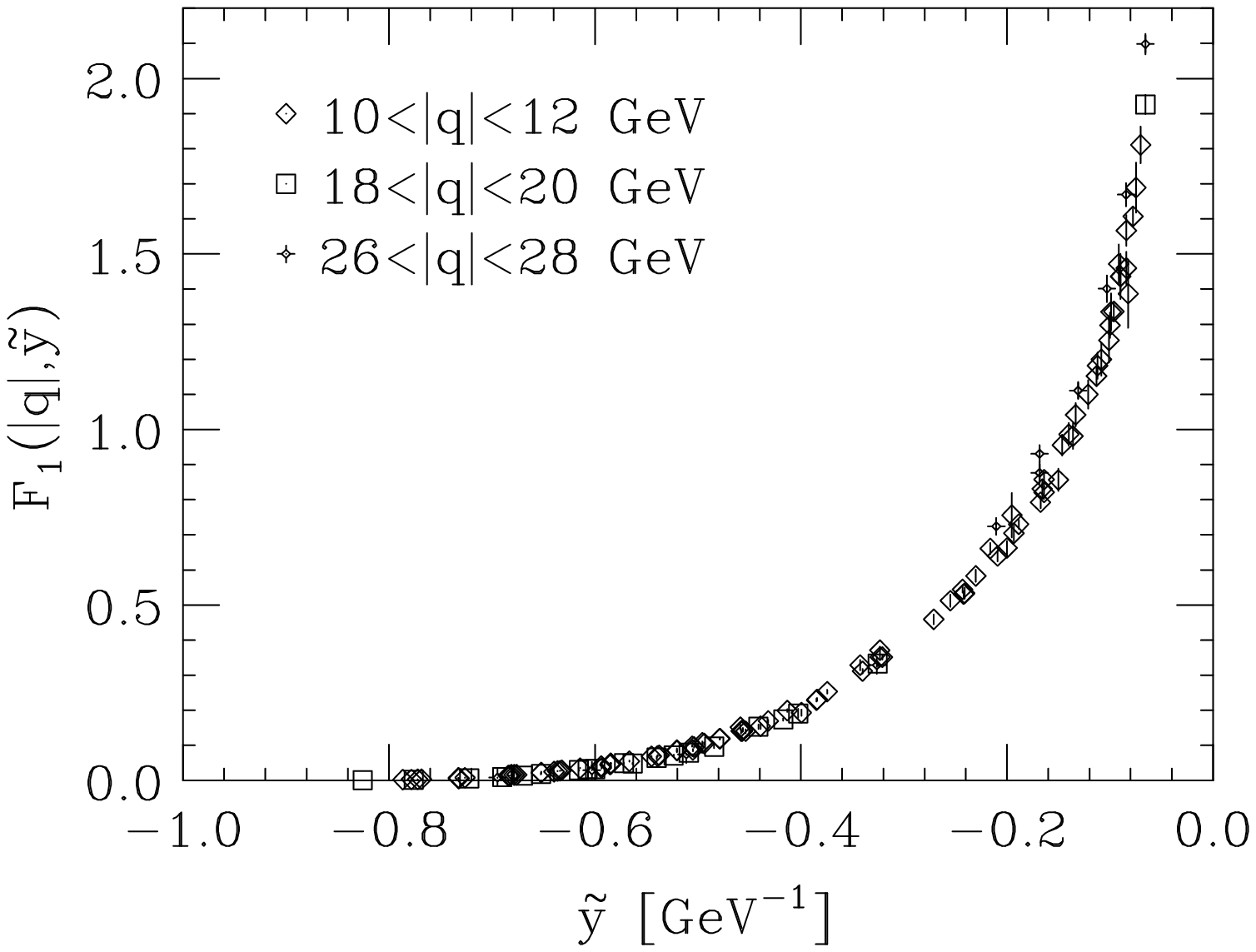,
angle=000,width=10.cm,height=09.0cm}}
\vspace*{.2in}
\caption{
$\yt$-dependence of the structure function $F_1 = M W_1$ 
obtained from the data of refs.\protect\ci{SLAC,NMC,BCDMS}.
The data sets are labelled according to the values of $\magq$.
}
\label{DIS:6}
\end{figure}

\begin{figure}
\vspace*{1.in}
\centerline
{\epsfig{figure=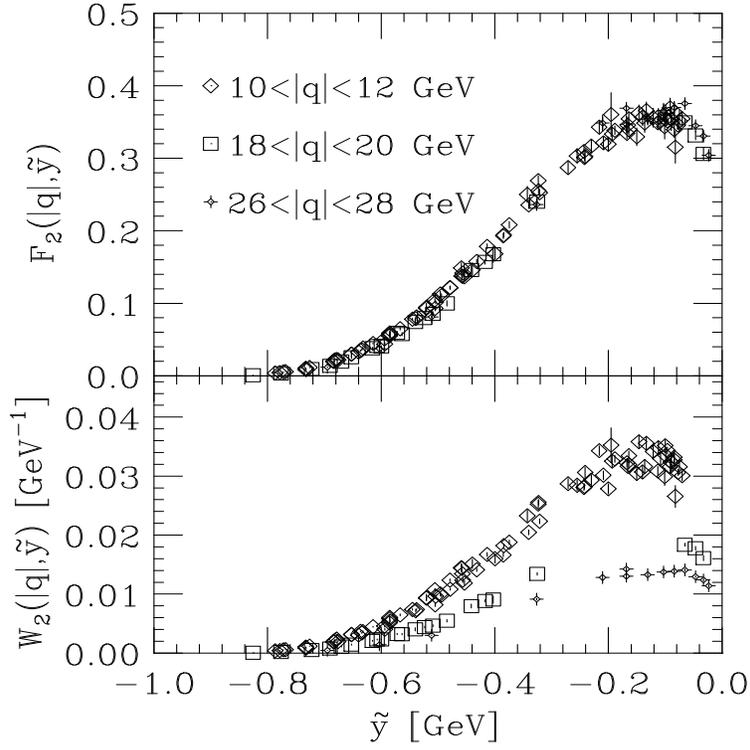,
angle=000,width=10.cm,height=10.0cm}}
\vspace*{.2in}
\caption{
$\yt$-dependence of $W_2$ (lower panel) and $F_2 = \nu W_2$
(upper panel) obtained from the data of refs.\protect\ci{SLAC,NMC,BCDMS}.
The data sets are labelled according to the values of $\magq$.
}
\label{DIS:7}
\end{figure}

As $F_1$ and $F_2$ are known to scale in $x$ and the Bjorken variable is 
nearly proportional to 
$\yt$, they obviously scale in $\yt$ as well. Figs. \ref{DIS:6} and \ref{DIS:7}
show that $F_1$ and $F_2$ do indeed exhibit $\yt$-scaling at fixed 
$\magq$, whereas $W_2$ does not. Note that the data plotted in 
Figs. \ref{DIS:6} and \ref{DIS:7} span a wide range of $Q^2$, extending from 
1.7 to 37.5 GeV$^2$.

\begin{figure}
\vspace*{1.in}
\centerline
{\epsfig{figure=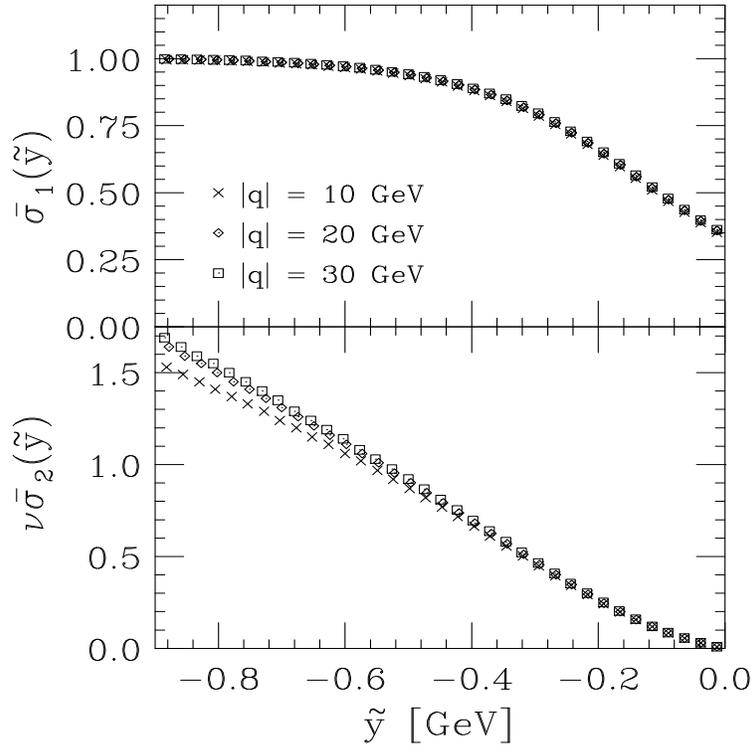,
angle=000,width=10.cm,height=10.0cm}}
\vspace*{.2in}
\caption{
Dependence of ${\overline \sigma}_1$ (upper panel) and
$\nu {\overline \sigma}_2$ (lower panel), 
given by  Eqs.(\protect\ref{sig:1})-(\protect\ref{sig:2}) and
(\protect\ref{w1:Q})-(\protect\ref{w2:Q}),  upon the momentum 
transfer $\magq$.
Crosses, diamonds and squares correspond to $\magq =$ 10, 20 and 30 GeV,
respectively. All calculations have been carried using $m =$ 300 MeV and
$B_0 =$ 200 MeV.
}
\label{DIS:4}
\end{figure}

It is very important to realize that, while $\yt$ essentially coincides 
with $-Mx$, 
the scaling function $F(\yt)$ cannot be identified with either $F_1$ or $F_2$. 
The structure of $F(\yt)$ is entirely dictated by the internal dynamics of
the target, whereas both 
$F_1 =  M \overline{\sigma}_1 F$ and $F_2 = \nu \overline{\sigma}_2 F$ contain 
part of the cross section describing the elecron-constituent interaction. 
The shape of $F_1$ and $F_2$ is in fact strongly affected by the presence of 
the electromagnetic cross section. For example, at $\yt = 0$ gauge invariance 
requires $F_2$ to vanish, whereas $F_1$ must be proportional to the 
photoabsorption cross section. 
The $\yt$-scaling behavior displayed by $F_1$ and $F_2$ is a consequence of the 
fact that, besides the proton response $F$, ${\overline \sigma}_1$ and 
$\nu {\overline \sigma}_2$ also scale. This feature is illustrated 
in Fig.\ref{DIS:4}, showing that ${\overline \sigma}_1$ and 
$\nu {\overline \sigma}_2$, plotted as a function of $\yt$, do not depend 
on $\magq$ for $\magq > 10$ GeV.

\begin{figure}
\vspace*{1.in}
\centerline
{\epsfig{figure=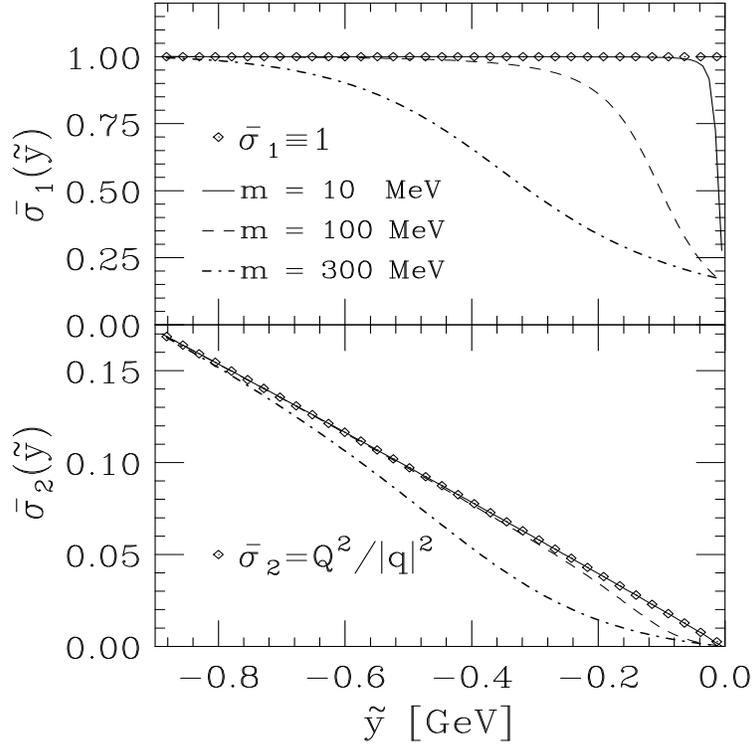,
angle=000,width=10.cm,height=10.0cm}}
\vspace*{.2in}
\caption{
Dependence of the elementary electron-constituent cross sections
${\overline \sigma}_1$ (upper panel) and ${\overline \sigma}_2$ (lower panel), 
given by  Eqs.(\protect\ref{sig:1})-(\protect\ref{sig:2}) and
(\protect\ref{w1:Q})-(\protect\ref{w2:Q}), 
upon the constituent mass. The solid, dashed and dot-dash lines correspond to 
$m =$ 10, 100 and 300 MeV, respectively, whereas the diamonds show the 
$m \rightarrow 0$ limit. All calculations have been carried at $\magq =$ 10 GeV
and using $B_0 =$ 0. 
}
\label{DIS:5}
\end{figure}

The different pictures emerging from $x$- and $\yt$-scaling analyses can only 
be reconciled making the standard assumption of parton model that the 
constituent 
mass be negligibly small. As shown in fig. \ref{DIS:5}, as $m \rightarrow 0$ 
\beq
{\overline \sigma}_1 \rightarrow 1 \ \ \  , 
\ \ \ {\overline \sigma}_2 \rightarrow  \frac{Q^2}{\magq^2}
\eeq
implying in turn that, in the Bjorken limit, 
\beq
F_1 = MF(\yt) \ \ \ , \ \ \ F_2 =  \nu \frac{Q^2}{\magq^2}\  F(\yt) = 2xF_1\ ,
\eeq
and the standard picture of DIS is recovered. However, 
it has to be emphasized 
that, although in textbook derivations the requirement $m \sim 0$ is 
often introduced as a necessary condition for scaling in DIS, the present 
analysis shows that scaling occurs irrespective of the constituent mass.

\section{Implications of the proposed analysis}

The differences between $\yt$-scaling analysis and the standard
$x$-scaling analysis, based on the parton model, have several interesting
implications for both the theoretical description of DIS and the
interpretation of experimental data.

The fact that the data appear to be compatible with the naive description 
of the proton in terms of massive bound constituents suggests that 
past attempts to describe DIS within the constituent quark model (CQM)
\ci{traini1,thomas} may have to be reconsidered. 

\begin{figure}
\vspace*{1.in}
\centerline
{\epsfig{figure=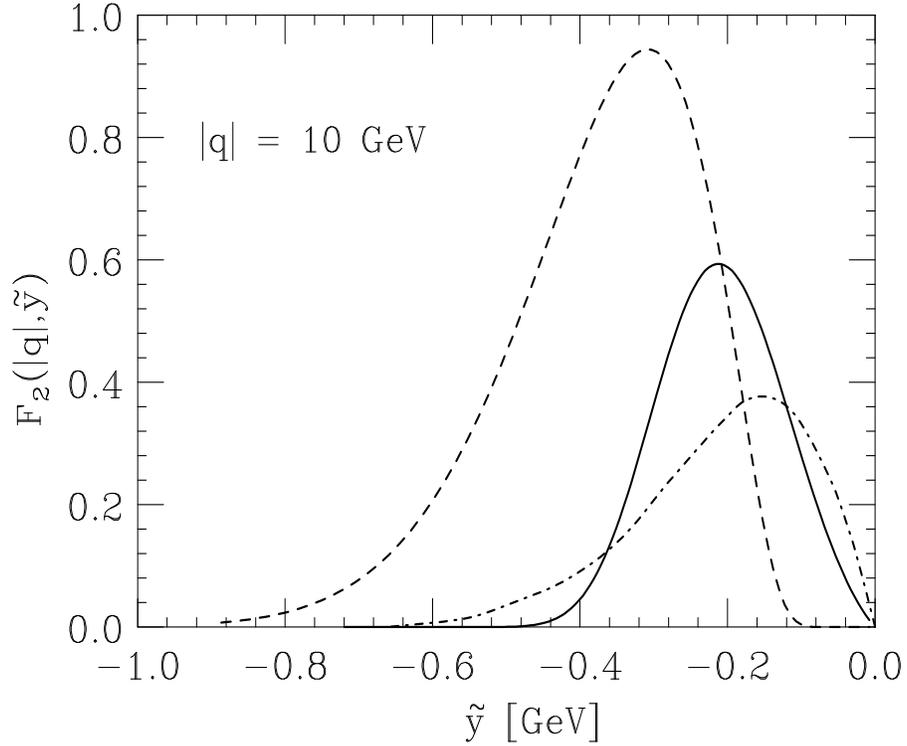,
angle=000,width=12.cm,height=10.0cm}}
\vspace*{.2in}
\caption{
Valence quarks contribution to the proton $F_2$, plotted as a function 
of $\yt$ at constant $\magq = 10\ {\rm GeV}$. The dashed and dash-dot lines 
show the results of the constituent quark model of ref.\protect\ci{traini1}
before and after QCD evolution until $Q^2 = 15\ {\rm GeV}^2$, respectively. 
The solid line shows $F_2$ obtained within the approach described in this 
paper, using the same quark mass and momentum distribution as in 
ref.\protect\ci{traini1} and $B_0 = 200\ {\rm MeV}$.
}
\label{DIS:8}
\end{figure}

Within the CQM, the valence quark distributions at low resolution scale 
$Q_0^2~\sim \mu_0^2~\sim (M/3)^2$, $u_v(x,Q_0^2)$ and $d_v(x,Q_0^2)$, are 
obtained from quark momentum 
distributions and used as a starting point for QCD evolution to the 
$Q^2$ scale relevant to DIS data. As an illustrative example of this procedure,
fig. \ref{DIS:8} shows the valence quarks contribution to the proton structure 
function $F_2$, evaluated in ref.\ci{traini1} using a harmonic 
oscillator quark momentum distribution, before (dashed line) and 
after (dot-dash line) QCD evolution until $Q^2 = 15\ {\rm GeV}^2$. The authors 
of ref.\ci{traini1} chose a constituent mass $m = M/3$ and adjusted the 
harmonic oscillator frequency in such a way as to reproduce the experimental 
value of the proton rms charge radius. 

The approach proposed in this paper provides a consistent alternative
framework to calculate the proton structure functions from CQM momentum 
distributions. For any given momentum distribution $n({\bf p})$, $F_2  = \nu W_2$ can 
in fact be obtained from Eq.(\ref{W12:IA}) using 
$P(p) = n({\bf p}) \delta(p_0 - \overline{p}_0)$, $\overline{p}_0$ being defined
in terms of the parameter $B_0$ through Eqs.(\ref{p0:bar}) and (\ref{recoiling:mass}).

The structure function $F_2$ obtained using the 
same constituent mass and momentum distribution as in ref.\ci{traini1} 
and $B_0 = 200\  {\rm MeV}$ is shown by the solid line in fig. \ref{DIS:8}.
Note that in the CQM calculation of the distributions $u_v(x,Q_0^2)$ and 
$d_v(x,Q_0^2)$, the constituent quarks are assumed to be on mass shell. 
As a consequence, $M_{\cal R} = M - m$, implying in turn $B_0=0$. 

\begin{figure}
\vspace*{1.in}
\centerline
{\epsfig{figure=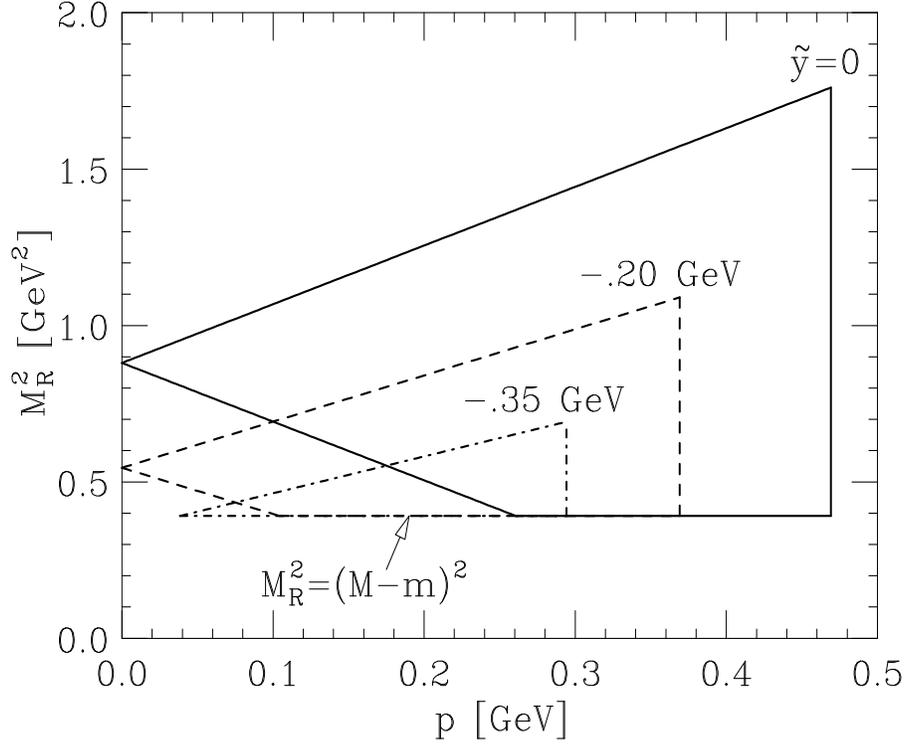,
angle=000,width=12.cm,height=10.0cm}}
\vspace*{.2in}
\caption{
Domains of the $(M_{\cal R}^2,|{\bf p}|)$ plane
relevant to the calculation of the proton structure functions
at $\magq = 10$ GeV and $\yt = 0$ (solid line), $-200$ MeV (dashed line) and 
$-350$ MeV (dot-dash line) from Eq.(\ref{W12:IA}). The results
have been obtained using $m = M/3$.
}
\label{DIS:9}
\end{figure}

Inclusion of  binding, described by $B_0 \neq 0$, leads to the 
appearance of strength in the region $\yt > -.10\ {\rm GeV}$, where the 
structure function of ref.\ci{traini1} vanishes. The mechanism responsible
for this feature can be readily understood with the help of fig. \ref{DIS:9}, 
showing the domains of the $(M_{\cal R}^2,|{\bf p}|)$ plane  
relevant to the calculation of the proton structure functions 
at $\magq = 10$ GeV and $\yt = 0, -200$ MeV and $-350$ MeV using Eq.(\ref{W12:IA}).

In the CQM, as the quark energy distribution is a $\delta$-function 
requiring 
$B_0=0$, the integration is carried out along the horizontal line
$M_{\cal R}^2 = (M - m)^2$, corresponding to the lower limit of the 
kinematically allowed domains. The maximum of the response is located
to $\yt=-m$, because in this case the lower limit of the momentum integration
is $|{\bf p}_{min}|=0$. Larger $\yt$ corresponds to larger 
 $|{\bf p}_{min}|$, leading in turn to a smaller response, as the 
quark momentum distribution is a rapidly decreasing function of $|{\bf p}|$.
For example, at $\yt=-.10$~GeV  $n(|{\bf p}_{min}|)/n(|{\bf p}|=0) \sim 1 \%$.
Setting $B_0 \neq 0$, which amounts to carrying out the integration along 
the horizontal line $M_{\cal R}^2 = (M - m + B_0)^2$, leads to a decrease
of $|{\bf p}_{min}|$ at $\yt > -m$, i.e. to an increase of the response. 
At $\yt=-.1$ GeV, one finds that $B_0 = 0$ and $200$ MeV 
correspond to $|{\bf p}_{min}| = .20$ and $.01$ GeV, respectively.

The fact that scaling in $\yt$ occurs irrespective of the values of $m$ and 
$B_0$, the resulting scaling function being roughly determined by the 
difference $m - B_0$, i.e.  by the mass of the spectator system 
$M_{\cal R}$, poses a serious problem for the extraction of $F(\yt)$
from the data and its interpretation. Even under the standard parton model 
assumption of negligibly small mass, allowing the constituent to be off mass 
shell leads to a shift of $F(\yt)$ towards larger $\yt$. 

In ref.\ci{bps} it has been pointed out that if binding effects are 
large, i.e. if the energy distribution of the constituents extends into 
the region of large $M_{\cal R}$, which dominates the response at $\yt \sim 0$ 
(see fig. \ref{DIS:9}), they may result in part of the response being pushed 
into the timelike region $\nu > \magq$, inaccessible to electron scattering. 
The occurrence of strength located in the timelike region is
a well known feature of the response of interacting many-body systems, 
simply reflecting the fact that the there is no {\it a priori} reason 
preventing the target from being left in a final state
with energy $E_n > E_0 + \magq$, where $E_0$ is the energy of the initial 
state. 
On the other hand, in the case of noninteracting systems the inequality 
$\nu < \magq$ is always satisfied 
and the response can be shown to vanish identically for $\yt > 0$.
This issue has been recently investigated in ref.\ci{mark1}, whose authors 
have shown that $\sim 10\ \%$ of the strength associated with the response of 
a confined relativistic particle is indeed located in the region 
of positive~$\yt$.

Obviously, the presence of strength in the timelike region
would make the experimental verification of sume rules involving the 
proton response, e.g. the Gross-Llewellyn Smith sum rule \ci{sumrule}, 
impossible. This problem has long been recognized in the analysis of
electron-nucleus scattering data. For example, integration of the measured 
charge response of iron at momentum transfer $|{\bf q}| = 570\  {\rm MeV}$
yields only $\sim 90 \%$ of the nuclear charge Z \ci{jourdan}.

\section{Conclusions}

The results described in this paper show that the scaling analysis of 
scattering 
processes off many-body systems, successfully employed to describe 
neutron-liquid helium and electron-nucleus scattering, can be extended 
to the case of DIS of electrons by protons.

The data exhibit a remarkable scaling behavior in the natural variable of
many-body theory, whose definition and interpretation
naturally emerge from the analysis of the scattering process in the target rest
frame. The onset of ${\widetilde y}$-scaling follows from the
assumptions underlying the IA picture. No further assumptions on the mass and 
binding energy of proton constituents are required.

While in the Bjorken limit $\yt$ essentially concides with $x$, up to a 
constant factor, the scaling function $F(\yt)$ cannot be trivially related 
to the structure functions $F_1(x)$ and $F_2(x)$, whose behavior is dictated
by both the internal proton dynamics and the electromagnetic 
electron-constituent interaction. 
The correspondence between $F(\yt)$ and $F_1(x)$ and $F_2(x)$ can 
only be established in the limit of vanishing constituent mass. 

The function $F(\yt)$, which can be identified with the target response 
in the scaling limit, turns out to be significantly affected by binding
effects, the position of its peak being mainly dictated by the mass of the 
spectator system $M_{{\cal R}}$.

The structure function $F_2$ calculated from Eq.(\ref{W12:IA}) appreciably 
differs from the prediction of the CQM at low resolution scale. 
Comparison between the solid and dashed curves of fig. \ref{DIS:9} shows that
the treatment of binding and mass according to the approach discussed in this 
paper produces an effect qualitatively similar to that of QCD evolution, shifting
strength towards larger $\yt$. The results displayed in fig. \ref{DIS:9} suggest 
that the role of QCD 
evolution should be reassessed, using the structure function represented by 
the solid line as a starting point.

Binding effects, resulting in a shift of $F(\yt)$ towards
larger values of $\yt$, can also push a fraction of the strength into the 
region of positive $\yt$, inaccessible to DIS experiments.  The proposed 
analysys, being carried out at fixed $\magq$, appears to be
best suited to address this feature, as moving along lines of constant $\magq$
one intersects the photon line $\nu = \magq$ and enters the timelike region, 
while 
moving along lines of constant $Q^2>0$, as generally done in the analysis
of DIS, the photon line is only asymptotically approached \ci{bps}.

Finally, it has to be pointed out that 
the observation of scaling {\it does not} necessarily imply the 
validity of the assumptions underlying the IA picture, i.e. dominance of
incoherent scattering and absence of FSI. 
It has been shown that, both in neutron-liquid             
helium scattering \ci{negele} and in electron-nucleus 
scattering \ci{omar}, as long as the cross section describing the interactions 
between 
target constituents does not depend upon $\magq$, FSI {\it do not} prevent the
occurrence of scaling. However, in presence of FSI the scaling function 
extracted from the data {\it can not} be related to the initial 
momentum and energy distribution of target constituents. 

The results of ref.\ci{mark2} indicate that sizable FSI effects in the 
response of a scalar relativistic particle confined by a linear potential
persist in the kinematical regime of large $\magq$, in which 
$\yt$-scaling is clearly observed. 
The role of FSI in DIS and their impact on the interpretation of the
measured proton strucure functions have been recently discussed in 
refs.\ci{sannino,sannino2}, whose authors conclude that the DIS cross section 
is indeed affected by final state rescatterings, and can not be simply related 
to the parton distributions in the initial state.

\begin{acknowledgments}
The author is deeply indebted to Vijay R. Pandharipande and 
Ingo Sick for many illuminating discussions on different issues related to 
the subject of this paper. 
The hospitality of the Theory Group at the Thomas Jefferson National 
Accelerator Facility and the partial support of the U.S. Department of Energy
during the completion of this work are also gratefully acknowledged.
\end{acknowledgments}

\appendix
\section{Target electromagnetic tensor within IA}

 The general expression of the target electromagnetic tensor
\beq
W^{\mu\nu} = \sum_n\  \int d^3p_n \langle 0 | J^\mu | n \rangle
 \langle n | J^\nu | 0 \rangle \delta^{(4)}(P_0 + q - P_n)
\label{A:1}
\eeq
drastically simplifies under the assumptions underlying 
the IA scheme, which lead to rewrite the target current and final 
state as (the index $i$ labels the struck constituent)
\beq
J^\mu \rightarrow \sum_i j_i^\mu 
\label{cont:curr}
\eeq
and
\beq
| n \rangle \ \rightarrow \  | {\bf p}^\prime,{\cal R} \rangle =  
|i({\bf p}^\prime) \rangle \otimes
|{\cal R}({\bf p}_{_R}) \rangle\ .
\label{fin:state}
\eeq
In the above equations $j_i^\mu$ is the current associated to a
target constituent, whereas the states $|i({\bf p}^\prime) \rangle$ and
$|{\cal R}({\bf p}_{_R}) \rangle$ describe the struck constituent and
the spectator system, respectively (see the discussion following 
eq.(\ref{energy:IA})). The structure of the IA final state can be exploited
to rewrite the sum appearing in Eq.(\ref{A:1}) replacing
\beq
\sum_{n} | n \rangle \langle n | \  \rightarrow \
 \int d^3p^\prime\ |{\bf p}^\prime  \rangle \langle {\bf p}^\prime |
 \sum_{{\cal R}} | {\cal R} \rangle \langle {\cal R} |\ . 
\label{sum:states}
\eeq

Substitution of Eqs.(\ref{cont:curr})-(\ref{sum:states}) into Eq.(\ref{A:1})
and insertion of complete sets of free particle states, satisfying 
\beq
\int d^3p\ | {\bf p} \rangle \langle {\bf p} | = I\ ,
\eeq
yields the expression
\be
\nonumber
W^{\mu\nu}  & = & \sum_i \sum_{\cal R}\  \int d^3p\ d^3p^\prime
\left| \langle 0 | {\bf p}, {\cal R} \rangle \right|^2 \ 
 \langle {\bf p} | j^\mu_i | {\bf p}^\prime \rangle
    \langle {\bf p}^\prime | j^\nu_i | {\bf p} \rangle \ \ \ \ \ \ \ 
\ \ \ \ \ \ \ \ \ \ \ \ \ \\
& & \ \ \ \ \ \ \ \ \ \ \ \ \ \ \ \ \ \ \ \times 
 \delta^{(3)}({\bf q} + {\bf p} - {\bf p}^\prime) \delta(\nu+E_0-E_{\cal R}
-E_{{{\bf p}^\prime}})\ ,
\ee
that can be further rewritten in a more compact form in terms of the 
constituent momentum end energy distribution $P(p)$, defined by 
Eq.(\ref{spec:fcn}). As a result we obtain
\beq
W^{\mu\nu} = \sum_i \  \int d^4p\ P(p)
 \int d^3p^\prime
\langle {\bf p} | j^\mu_i | {\bf p}^\prime \rangle
\langle {\bf p}^\prime | j^\nu_i | {\bf p} \rangle
 \delta^{(3)}({\bf q} + {\bf p} - {\bf p}^\prime)
\delta(\nu+p_0 -E_{{{\bf p}^\prime}})\ .
\label{A:7}
\eeq
Finally, we can substitute in the above equation the definition of the 
electromagnetic tensor 
corresponding to scattering at four-momentum tranfer 
$q \equiv(\nu,{\bf q})$
off a {\it free} constituent carrying momentum ${\bf p}$:
\beq
w^{\mu\nu}_i =
 \int d^3p^\prime
\langle {\bf p} | j^\nu_i | {\bf p}^\prime \rangle
\langle {\bf p}^\prime | j^\nu_i | {\bf p} \rangle
 \delta^{(3)}({\bf q} + {\bf p} - {\bf p}^\prime)
\delta(\nu+E_{{\bf p}}-E_{{{\bf p}^\prime}})\ .
\label{A:8}
\eeq
Comparison of Eqs.(\ref{A:7}) and (\ref{A:8}) shows that 
$W^{\mu\nu}$ can be rewritten in the form
\beq
W^{\mu\nu} = \sum_i \  \int d^4p\ P(p)\ 
w^{\mu\nu}_i({\widetilde p},{\widetilde q})\ ,
\label{A:9}
\eeq
where ${\widetilde p} \equiv(E_{{\bf p}},{\bf p})$ and 
${\widetilde q} \equiv ({\widetilde \nu},{\bf q})$, with 
${\widetilde \nu} = E_{{{\bf p}+ {\bf q}}} - E_{{\bf p}} = 
\nu + p_0 -  E_{{\bf p}}$.


\section{Restoration of gauge invariance}

The definition of the tensor $W^{\mu\nu}$ given by Eq.(\ref{A:9}) obviously 
leads to breaking gauge invariance, as
\beq
q_\mu w^{\mu\nu}_i({\widetilde p},{\widetilde q}) \neq 0
\eeq
by construction. 

The prescription proposed in ref.\ci{defo} to circumvent this problem 
amounts to using IA only for the transverse and time components
of the current $j_i^\mu$. Denoting by ${\widetilde w}_{\mu \nu}$ the 
electromagnetic 
tensor of ref.\ci{defo} and choosing the $z$-axis along the direction of
${\bf q}$ we can then write
\beq
{\widetilde w}_i^{\mu\nu}(p,q) =
w_i^{\mu\nu}({\widetilde p},{\widetilde q})\ \ \ \ , \ \ \ \ \ \mu, \nu \ne 3\ .
\eeq
The longitudinal current $j_i^3$ is defined in terms of the time component 
$j_i^0$ in such a way as to satisfy the 
continuity equation. As a result, the corresponding elements of the tensor 
${\widetilde w}_i^{\mu\nu}$ read
\beq
{\widetilde w}_i^{\mu 3}(p,q) =
\left( \frac{\nu}{|{\bf q}|} \right)
w_i^{\mu 0}({\widetilde p},{\widetilde q})
\eeq
and
\beq
{\widetilde w}_i^{3 3}(p,q) =
\left( \frac{\nu}{|{\bf q}|} \right)^2
w_i^{0 0}({\widetilde p},{\widetilde q})\ .
\end{equation}

The above procedure to reconcile the IA scheme with gauge invariance is
manifestly non unique. However, it has to be emphasized that, since 
according to ref.\ci{defo} the current satisfies the relation 
${\widetilde j^3} = (\nu / \magq ) j^0$, while 
the IA current satisfies $j^3 = ({\widetilde \nu} / \magq ) j^0$, the 
difference
\beq
{\widetilde j}^3 - j^3 = \frac{ \nu - {\widetilde \nu} }{\magq}
 = \frac{ E_{{\bf p}} - p_0 }{\magq}\ ,
\eeq
providing a measure of the violatation of gauge invariance,
becomes vanshingly small in the $\magq \rightarrow \infty$ limit.


\clearpage

\clearpage

\clearpage

\clearpage

\clearpage

\clearpage

\clearpage

\clearpage

\clearpage

\end{document}